\documentclass[prb,aps,twocolumn,floatfix,superscriptaddress,letterpaper,longbibliography]{revtex4-2}
\usepackage{bm}
\usepackage[english]{babel}
\usepackage{amsmath,amsfonts}
\usepackage{amsbsy}
\usepackage[colorlinks=true,linkcolor=blue,urlcolor=blue,citecolor=blue,breaklinks=true]{hyperref}
\usepackage[utf8]{inputenc}
\DeclareUnicodeCharacter{03BC}{\mbox{$\mu$}}
\DeclareUnicodeCharacter{2009}{ }

\begin{document}

\title{Viscous dissipation in a gas of one-dimensional fermions with generic dispersion}

\author{Wade DeGottardi}

\affiliation{Texas Tech University, Department of Physics and Astronomy, Lubbock, Texas 79409, USA }

\author{K. A. Matveev}

\affiliation{Materials Science Division, Argonne National Laboratory,
  Argonne, Illinois 60439, USA}

\date{\today}

\begin{abstract}
A well-known feature of the classical monoatomic gas is that its bulk viscosity is strongly suppressed because the single-particle dispersion is quadratic. On the other hand, in condensed matter systems the effective single-particle dispersion is altered by lattice effects and interactions. In this work, we study the bulk viscosity of one-dimensional Fermi gases with generic energy-momentum dispersion relations. As an application, viscous dissipation arising from lattice effects is analyzed for the tight-binding model. In addition, we investigate how weak interactions affect the bulk viscosity. Finally, we discuss viscous dissipation in the regime in which the Fermi gas is not fully equilibrated, as can occur when the system is driven at frequencies that exceed the rate of fermion backscattering. In this case, the Fermi gas is described by three bulk viscosities, which we obtain for a generic single-particle dispersion.
\end{abstract}

\maketitle

\section{Introduction}

The behavior of fluids perturbed from mechanical and thermal equilibrium is commonly described by classical hydrodynamics. In this approach, dissipative effects are accounted for by phenomenological parameters known as transport coefficients~\cite{landau_fluid_1987}. These quantities, which include the viscosity and thermal conductivity, characterize dissipation that arises due to small perturbations. In particular,
viscosity quantifies the dissipation that arises from a non-uniform velocity of the fluid. Of late, there has been a surge of interest in the application of hydrodynamics to low-dimensional gases and liquids of fermions~\cite{joseph_observation_2011,bandurin_negative_2016,crossno_observation_2016,moll_evidence_2016,levitov_electron_2016,lucas_transport_2016,guo_higher-than-ballistic_2017}.
Viscosity plays a determinative role in a variety of non-equilibrium behaviors displayed by such systems, including transport properties of quantum wires~\cite{andreev_hydrodynamic_2011} as well as the relaxation of the collective modes in cold atomic gases~\cite{vogt_scale_2012}.

For three-dimensional systems, viscous effects are captured by two transport coefficients, the shear and bulk viscosities. In one dimension shear is not defined, and thus the relevant transport coefficient is the bulk viscosity. In the present work, we study the bulk viscosity of a one-dimensional (1D) Fermi gas with an arbitrary single-particle dispersion. To appreciate the central role that the dispersion plays in viscous effects, consider the classical monoatomic gas. As is well-known, its bulk viscosity is suppressed because the single-particle dispersion is quadratic~\cite{lifshitz_physical_1981}. On the other hand, in condensed matter systems the dispersion is altered by lattice effects. These effects, even if small, are expected to significantly enhance the bulk viscosity $\zeta$. One of the key results of the present work is an expression for $\zeta$ for a generic dispersion.

Interactions play a crucial role in viscous dissipation, as they cause the collisions responsible for the relaxation of a system to equilibrium. Interactions also alter the form of the dispersion. Typically, dispersion relations are discussed in the context of single-particle dynamics. However, an \emph{effective} single particle dispersion that accounts for interactions perturbatively can be defined. We derive this effective dispersion and use it to obtain an expression for the bulk viscosity. In an experiment, both interactions and lattice effects can be weak. We thus consider the competition between these effects. In particular, this physics is investigated in the case of a dilute Fermi gas using a tight-binding dispersion as an example.

The application of hydrodynamics requires that the system be close to equilibrium. A unique feature of the relaxation of 1D systems is that not all relaxation processes have rates of the same order of magnitude. Typical thermal excitations relax at a rate that scales as a power of temperature. In contrast, fermion backscattering occurs at a rate $1/\tau_b$, which is exponentially suppressed at low temperatures~\cite{lunde_three-particle_2007,micklitz_transport_2010,matveev_scattering_2012}. When a 1D system is driven at a frequency $\omega \gg 1/\tau_b$, backscattering is frozen out, and thus the system cannot fully equilibrate. As long as $\omega$ is less than all other relaxation rates, the behavior of this system is described by two-fluid hydrodynamics~\cite{matveev_second_2017,matveev_propagation_2018,matveev_two-fluid_2019}, analogous to the well-known theory of superfluid He-4~\cite{landau_fluid_1987,khalatnikov_introduction_2000}. As in the superfluid, accounting for viscous dissipation in the driven Fermi gas requires three bulk viscosities~\cite{matveev_propagation_2018}. We calculate these three transport coefficients for an arbitrary dispersion.

The paper is organized as follows. Section~\ref{sec:boltzmann} introduces our approach for the calculation of the bulk viscosity for a generic dispersion and applies it to the tight-binding dispersion. In Sec.~\ref{sec:finite_frequencies}, we consider the viscous coefficients for a gas in the two-fluid regime. We study the effect of weak interactions on viscous dissipation in Sec.~\ref{sec:interactions}. In Sec.~\ref{sec:scaleinvariance}, we discuss the suppression of the bulk viscosity for specific dispersions. Finally, in Sec.~\ref{sec:discconcl}, we highlight the implications of our results.

\section{Bulk Viscosity and the Boltzmann Equation}

\label{sec:boltzmann}

In this section, we derive the bulk viscosity for a general dispersion $\varepsilon_p$ (Sec.~\ref{sec:boltzmannA}) and apply this result to the tight-binding model (Sec.~\ref{sec:tight-binding}). Viscous dissipation arises when the velocity of the gas is not uniform. In the usual case of a Galilean invariant system, the velocity of an element of the fluid is that of its center of mass. For a generic dispersion, however, mass is not defined. We thus use an alternative definition of velocity based on the equilibrium distribution function. In a translationally invariant Fermi gas, not only particle number and energy but also momentum is conserved, and the equilibrium distribution function takes the form
\begin{equation}
n_p^{(0)} = \frac{1}{ e^{\left(\varepsilon_p - u p - \mu\right)/T} + 1}.
\label{eq:neq}
\end{equation}
Here $p$ is the momentum of the state, while $T$, $u$, and $\mu$ control the average energy, momentum, and particle number. The parameters $T$ and $\mu$ are the temperature and chemical potential of the gas, respectively. The parameter $u$ has the dimension of velocity. In the case a Galilean invariant system, $\varepsilon_p = p^2/2m$, it coincides with the center of mass velocity of the gas. For generic $\varepsilon_p$, we take $u$ appearing in Eq.~(\ref{eq:neq}) as the definition of the velocity of the gas. In the following we assume for simplicity that $\varepsilon_p$ is even in $p$ and monotonically increasing for positive $p$. At $T \to 0$ such a gas only has a single pair of Fermi points.

A gas with a spatially varying velocity $u(x)$ is not in equilibrium. For infinitesimal $\partial_x u$, the distribution function is given by
\begin{equation}
n_p = n_p^{(0)} + \delta n_p,
\label{eq:ansatz}
\end{equation}
where $\delta n_p$ is the infinitesimal dissipative part of $n_p$. While the perturbation $\partial_x u \neq 0$ drives the system out of equilibrium, collisions between particles caused by weak interactions tend to restore it, resulting in a non-zero rate of change of the distribution function, which we denote by $\dot{n}_p$. The power dissipated per unit length is given by
\begin{equation}
w = - \nu T \int_{-\infty}^{\infty} \frac{dp}{h} \frac{\dot{n}_p \, \delta n_p}{n_p^{(0)} \left( 1 - n_p^{(0)}  \right)} ,
\label{eq:w2}
\end{equation}
where $h$ is Planck's constant and $\nu$ is the degeneracy associated with spin, with $\nu = 2S+1$ for a gas of spin-$S$ fermions. The expression for $w$ was derived~\cite{matveev_viscous_2017,degottardi_viscous_2020} by evaluating $T \dot{s}$, where the entropy density $s$ was expressed in terms of the Fermi occupation numbers. Both $\delta n_p$ and $\dot{n}_p$ vanish for a system in equilibrium and therefore must be proportional to $\partial_x u$. Thus, the power dissipated per unit length (\ref{eq:w2}) has the form
\begin{equation}
w = \zeta \left( \partial_x u \right)^2.
\label{eq:zetadef}
\end{equation}
The proportionality constant $\zeta$ is the bulk viscosity. This definition is directly analogous to the definition of $\zeta$ for a Galilean-invariant fluid~\cite{landau_fluid_1987}. In the absence of Galilean invariance, $\zeta$ is a function of $u$. From now on, we limit ourselves to the study of the bulk viscosity at $u = 0$.

\subsection{The Boltzmann equation approach}

\label{sec:boltzmannA}

We now calculate $\zeta$ using Eqs.~(\ref{eq:w2}) and (\ref{eq:zetadef}). This requires us to obtain $\dot{n}_p$ and $\delta n_p$. In the Boltzmann equation formalism, $\dot{n}_p$ can be expressed in two ways:
\begin{eqnarray}
\dot{n}_p &=& \partial_t n_p + \left( \partial_p \varepsilon_p \right) \partial_x n_p - \left( \partial_x \varepsilon_p \right) \partial_p n_p, \label{eq:boltzmann1} \\
\dot{n}_p &=& I [ n_p ],
\label{eq:boltzmann2}
\end{eqnarray}
where $I[n_p]$ is the collision integral describing the relaxation to equilibrium. The standard Boltzmann equation~\cite{lifshitz_physical_1981} is obtained by equating these two expressions for $\dot{n}_p$. Interactions are responsible for the collisions between the particles. They also alter the effective dispersion $\varepsilon_p$ appearing in Eq.~(\ref{eq:boltzmann1}), an effect that will be considered in Sec.~\ref{sec:interactions}. In this section, the third term on the right hand side of Eq.~(\ref{eq:boltzmann1}) vanishes because $\partial_x \varepsilon_p = 0$. This term will play a role later in the paper.

A system with a velocity gradient is either expanding or contracting. As a result, the temperature $T$ and chemical potential $\mu$ depend on time. This is in contrast to the calculation of thermal conductivity, which can be obtained from a steady-state solution of the Boltzmann equation~\cite{matveev_thermal_2019}.
Substituting the expression (\ref{eq:neq}) for $n_p^{(0)}$ into Eq.~(\ref{eq:boltzmann1}) and allowing for the dependences $u(x)$, $T(t)$, and $\mu(t)$, we find
\begin{equation}
\dot{n}_p = \frac{1}{T} n_p^{(0)} \left(1 - n_p^{(0)} \right) \left[ \frac{ \varepsilon_p - \mu}{T} \partial_t T + \partial_t \mu + p \partial_p \varepsilon_p \partial_x u \right].
\label{eq:boltzmann3}
\end{equation}
This expression can be written more compactly as
\begin{equation}
\dot{n}_p = \frac{1}{T} n_p^{(0)} \left( 1 - n_p^{(0)}\right) \Upsilon(\xi),
\label{eq:leftboltz4}
\end{equation}
where $\xi = \varepsilon_p - \mu$ and
\begin{equation}
\Upsilon(\xi) =  \partial_t \mu + \frac{\partial_t T}{T} \xi  +  \partial_x u \frac{p(\mu+\xi)}{p'(\mu+\xi)}.
\label{eq:upsilon}
\end{equation}
The function $p(\varepsilon)$ is the inverse of $\varepsilon_p$ for positive $p$, and $p'(\varepsilon)$ is its derivative.

Next, we use conservation laws to express $\partial_t \mu$ and $\partial_t T$ in terms of $\partial_x u$. The conservation of particle number, momentum, and energy can be expressed as
\begin{eqnarray}
\int_{-\infty}^{\infty} \frac{dp}{h} \dot{n}_p X_p = 0,
\label{eq:conservation}
\end{eqnarray}
for $X_p = 1, p,$ and $\varepsilon_p$, respectively. While conservation of momentum is trivially satisfied given that the right-hand side of Eq.~(\ref{eq:boltzmann3}) is even in $p$, the conservation of particle number and energy gives two linear relations involving the infinitesimal quantities $\partial_t T$, $\partial_t \mu$, and $\partial_x u$, thus allowing us to express $\dot{n}_p$ in terms of $\partial_x u$ alone. Working to leading order in $T$ with $\xi \sim T$, we obtain
\begin{equation}
\dot{n}_p = \frac{g_p}{2T} \Upsilon''(0) \phi_p,
\label{eq:np}
\end{equation}
where
\begin{equation}
\phi_p = g_p \left( \xi^2 - \frac{\pi^2 T^2}{3} \right),
\label{eq:phi}
\end{equation}
and
\begin{equation}
g_p = \sqrt{n_p^{(0)} \left( 1 - n_p^{(0)} \right)} = \frac{1}{2 \cosh{\frac{\xi}{2T}}}.
\end{equation}
The result for $\dot n_p$ given by Eqs.~(\ref{eq:np}) and (\ref{eq:phi}) holds for any $\Upsilon(\xi)$ in Eq.~(\ref{eq:leftboltz4}), as long as the second derivative $\Upsilon''(0)$ is well defined. For $\Upsilon(\xi)$ given by Eq.~(\ref{eq:upsilon}), we have
\begin{equation}
\Upsilon''(0) = - \chi^{\phantom\dagger}_0 \partial_x u,
\label{eq:upsilonzero}
\end{equation}
where
\begin{equation}
\chi^{\phantom\dagger}_0  = - \left. \left( \frac{p(\varepsilon)}{p'(\varepsilon)} \right)'' \right\vert_{\varepsilon = \mu}  = - \left. \frac{1}{\varepsilon_p'} \left( \frac{p \, \varepsilon_p''}{\varepsilon_p'} \right)' \right\vert_{p = p_F}.
\label{eq:chi}
\end{equation}
Here the Fermi momentum $p_F$ is defined by $\varepsilon_{p_F} = \mu$. Of the two equivalent forms for $\chi^{\phantom\dagger}_0$ given in Eq.~(\ref{eq:chi}), the second is more convenient for a given dispersion. The sign in the definition (\ref{eq:chi}) of $\chi_0$ is chosen so that $\chi_0$ is positive for the tight-binding model (Sec. \ref{sec:tight-binding}).

The final ingredient necessary to calculate $\zeta$ using Eqs.~(\ref{eq:w2}) and (\ref{eq:zetadef}) is $\delta n_p$. As we now demonstrate, $\delta n_p$ may be obtained from Eq.~(\ref{eq:boltzmann2}) by linearizing the collision integral and inverting it. To do so, it is convenient to symmetrize the collision integral by introducing the function $x_p$ defined by $\delta n_p = g_p x_p$~\cite{buot_relaxation_1972}. For infinitesimal $x_p$, its rate of change is
\begin{equation}
\dot{x}_p = - \hat{\Gamma} x_p,
\label{eq:gamma}
\end{equation}
where the linear operator $\hat{\Gamma}$ is given by
\begin{equation}
\hat{\Gamma} x_p = - \frac{1}{g_p} \left. \frac{d}{d s} I \left[ n_p^{(0)} + s g_p x_p \right] \right|_{s = 0}.
\label{eq:linearized}
\end{equation}
(The parameter $s$ has been used to linearize $I$ in $x_p$.) The operator $\hat{\Gamma}$ is symmetric and thus its eigenvalues are real. Furthermore, the eigenvalues must be nonnegative in order for $n_p^{(0)}$ to represent a stable equilibrium.

We now formally express $x_p$ by inverting the linear operator $\hat{\Gamma}$ appearing in Eq.~(\ref{eq:gamma}). This gives
\begin{equation}
x_p = \frac{\chi^{\phantom\dagger}_0}{2T} \left( \partial_x u \right) \hat{\Gamma}^{-1} \phi_p,
\label{eq:xpinverse}
\end{equation}
where we have used Eqs.~(\ref{eq:np}) and (\ref{eq:upsilonzero})~\footnote{\label{foot:zero} The operator $\hat{\Gamma}^{-1}$ appearing in Eq.~(\ref{eq:xpinverse}) is well-defined as long as it acts on the subspace of eigenvectors with strictly non-zero eigenvalues. Indeed, $\phi_p$ belongs to this subspace because our procedure of using the conservation laws to express $\dot{n}_p$ in terms of $\partial_x u$ alone ensures that $\phi_p$ is orthogonal to the zero modes of $\hat{\Gamma}$. Thus, the operator $\hat{\Gamma}^{-1}$ in Eq.~(\ref{eq:xpinverse}) is well-defined.}. Then, Eq.~(\ref{eq:w2}) may be written in the compact form
\begin{equation}
w = \frac{\chi_0^2}{4 T} \langle \phi | \hat{\Gamma}^{-1} | \phi \rangle \left( \partial_x u \right)^2,
\label{eq:w3}
\end{equation}
where the inner product is defined by
\begin{equation}
\langle a | b \rangle = \nu \int_{-\infty}^{\infty} \frac{dp}{h} a_p b_p,
\end{equation}
for generic functions $a_p$ and $b_p$. In particular, substitution of $\phi_p$ given by Eq.~(\ref{eq:phi}) for both $a_p$ and $b_p$ yields
\begin{equation}
\langle \phi | \phi \rangle = \frac{16 \pi^3 \nu T^5}{45 \hbar v_F},
\label{eq:phiinner}
\end{equation}
at $T \ll E_F$. (Here the Fermi energy $E_F$ $ = \varepsilon_{p_F} - \varepsilon_0$).
Using Eqs.~(\ref{eq:zetadef}), (\ref{eq:w3}), and (\ref{eq:phiinner}), we obtain for the bulk viscosity
\begin{equation}
\zeta = \frac{4 \pi^3 \nu \chi_0^2 T^4 \tau }{45 \hbar v_F},
\label{eq:zeta}
\end{equation}
where the effective relaxation time $\tau$ is defined by
\begin{equation}
\tau = \frac{\langle \phi | \hat{\Gamma}^{-1} | \phi \rangle}{\langle \phi | \phi \rangle}.
\label{eq:tau}
\end{equation}
This expression for $\tau$ is the average of the inverse decay rates of the eigenmodes of the collision integral (\ref{eq:linearized}) weighted by their overlap with $\phi_p$.

The spectra of decay rates for 1D systems have been studied in a number of different cases~\cite{matveev_thermal_2019,matveev_relaxation_2020,matveev_scattering_2012,degottardi_equilibration_2019}. At low temperatures, the relaxation spectrum of a 1D Fermi gas exhibits two disparate rates. Fermionic backscattering occurs at a rate $1/\tau_b$, which is exponentially small at low temperatures~\cite{lunde_three-particle_2007,levchenko_transport_2010,micklitz_transport_2010,matveev_scattering_2012,matveev_thermal_2019}. All other relevant processes are comparatively fast, with rates that scale as a power of $T$. Importantly, backscattering is associated with a perturbation $x_p$ that is odd in $p$, while $\phi_p$ that appears in the definition (\ref{eq:tau}) is even. Therefore, only the fast modes contribute to Eq.~(\ref{eq:tau}), and the relaxation time $\tau$ scales as a power of temperature~\footnote{In contrast, the thermal conductivity is the response
to a nonzero gradient of temperature which, unlike $\partial_x u$, is odd
with respect to inversion. As a result, it is controlled by the exponentially long backscattering time $\tau_b$~\cite{matveev_relaxation_2020}.}.

For the quadratic dispersion $\varepsilon_p = p^2/2m$, Eq.~(\ref{eq:chi}) yields $\chi^{\phantom\dagger}_0 = 0$, and we recover the well-known result that the bulk viscosity is suppressed in this case~\cite{lifshitz_physical_1981}. We stop short of asserting that $\zeta$ given by Eq.~(\ref{eq:zeta}) vanishes since interactions can alter $\varepsilon_p$, as will be discussed in Secs.~\ref{sec:interactions} and \ref{sec:discconcl}.

In addition to the case of the quadratic dispersion, $\chi^{\phantom\dagger}_0$ also vanishes for the ultrarelativistic dispersion, $\varepsilon_p = c |p|$, cf.~Ref.~\cite{lifshitz_physical_1981}. We are thus led to ask what is the most general form of $\varepsilon_p$ for which $\chi^{\phantom\dagger}_0$ vanishes for any density. To answer this question, we set the second form of $\chi^{\phantom\dagger}_0$ given in Eq.~(\ref{eq:chi}) equal to zero for all $p_F$. The solution to the resultant third order differential equation gives a general dispersion of the form~\footnote{\label{foot:special} The condition that $\varepsilon_p$ is even necessitates the use of the absolute value in Eqs.~(\ref{eq:special_dispersion1}) and (\ref{eq:exp}). Additionally, the condition that $\varepsilon_p$ is monotonically increasing for positive $p$ requires that $A B > 0$.}
\begin{equation}
\varepsilon_p = A |p|^B + C.
\label{eq:special_dispersion1}
\end{equation}
This expression has as special cases the quadratic ($B = 2$) and ultra-relativistic ($B = 1$) dispersions. The physics of the vanishing of $\chi^{\phantom\dagger}_0$ is discussed in Sec.~\ref{sec:scaleinvariance}.

\subsection{Tight-binding dispersion}

\label{sec:tight-binding}

The bulk viscosity is sensitive to lattice effects through its dependence on $\chi^{\phantom\dagger}_0$, which in turn depends on $\varepsilon_p$. Here, we evaluate $\chi^{\phantom\dagger}_0$ for the tight-binding model in which the single particle dispersion relation is obtained by assuming that particles hop between neighboring sites. This model has been applied to a number of relevant systems, such as 1D fermions in optical lattices in the deep lattice regime~\cite{ibanez-azpiroz_tight-binding_2013}. In Sec.~\ref{sec:interactions}, we will apply the tight-binding dispersion to study the relative importance of viscous dissipation arising from lattice effects and interactions.

For a 1D lattice with spacing $a$, the tight-binding dispersion is
\begin{equation}
\varepsilon_p = \frac{D}{2} \left[ 1 - \cos \left( \frac{p a}{\hbar} \right) \right],
\label{eq:tight-binding-spectrum}
\end{equation}
where $D$ is the full bandwidth. This dispersion is characterized by the Fermi velocity
\begin{equation}
v_F = \varepsilon_{p_F}' = \frac{D a}{2 \hbar} \sin \left( \frac{p_F a}{\hbar} \right).
\label{eq:vF}
\end{equation}
The effects of the dispersion enter the bulk viscosity (\ref{eq:zeta}) through $v_F$ and $\chi^{\phantom\dagger}_0$. Substituting the tight-binding dispersion (\ref{eq:tight-binding-spectrum}) into Eq.~(\ref{eq:chi}), we obtain
\begin{equation}
\chi^{\phantom\dagger}_0 = \frac{1}{D} \frac{2 p_F a / \hbar - \sin \left( 2 p_F a / \hbar \right)}{\sin^3 \left( p_F a / \hbar \right)}.
\label{eq:chi_tight-binding}
\end{equation}
At $p_F \to 0$, we have $\chi^{\phantom\dagger}_0 \to 4 / 3 D$.

It is instructive to consider the behavior of $\chi^{\phantom\dagger}_0$ in the continuum limit. For a given particle density $n = \nu p_F / \pi \hbar$, the latter is achieved by requiring $a \rightarrow 0$ and $D \rightarrow \infty$, while $a^2 D$ is held fixed. In this limit, the dispersion (\ref{eq:tight-binding-spectrum}) approaches the form
\begin{equation}
\varepsilon_p = \frac{p^2}{2 m^\ast}, \quad
m^\ast = \frac{2\hbar^2}{D a^2}.
\label{eq:mstar}
\end{equation}
Since the dispersion (\ref{eq:mstar}) is quadratic, one expects $\chi^{\phantom\dagger}_0$ to vanish. Indeed, because $\chi_0 \rightarrow 4 / 3 D \propto a^2$ at $a \rightarrow 0$, we find that $\chi^{\phantom\dagger}_0$ does in fact vanish in the continuum limit.

\section{Viscous Response at Finite Frequencies}

\label{sec:finite_frequencies}

In our discussion of the relaxation time $\tau$, we noted that the bulk viscosity is not affected by backscattering. This conclusion, though derived for a time independent perturbation $\partial_x u \neq 0$, holds for the time-dependent case as long as the associated frequency obeys $\omega \ll 1/\tau_b$. For such frequencies, the system still comes to the equilibrium state~(\ref{eq:neq}), cf.~Ref.~\cite{matveev_second_2017}.
Since backscattering is the slowest relaxation process at low temperatures, there is a broad range of frequencies $1/\tau_b \ll \omega \ll 1/\tau$, for which backscattering is essentially frozen out, and the numbers of right and left movers are separately conserved~\cite{micklitz_transport_2010}. Because $\omega \ll 1/\tau$, at $T \ll E_F$ fast processes bring the system to \emph{partial} equilibrium~\cite{micklitz_transport_2010}, as described by the distribution function
\begin{equation}
n_p^{(0)} = \frac{1}{e^{\left( \varepsilon_p - u p - \mu - \textrm{sgn}(p) \delta \mu / 2 \right)/T}+1}.
\label{eq:dist2}
\end{equation}
The form of the partially equilibrated distribution function (\ref{eq:dist2}) is dictated by the fact that, at these frequencies, the left and right movers cannot come to diffusive equilibrium and thus are described by the distinct chemical potentials $\mu - \delta \mu/2$ and $\mu + \delta \mu / 2$, respectively.

We now consider the viscous effects that arise from gradients of $u$ and $\delta \mu$. It is necessary to establish the form of the dissipated power $w$ that generalizes Eq.~(\ref{eq:zetadef}). First, we observe that for $\delta \mu$ independent of position, $w$ must reduce to the form (\ref{eq:zetadef}) since the processes that underlie $\zeta$ are still operative for $\omega \ll 1/\tau$. From the argument that led to Eq.~(\ref{eq:zetadef}), it follows that for $\partial_x \delta \mu \neq 0$ and $\partial_x u = 0$, both $\dot{n}_p$ and $\delta n_p$ are proportional to $\partial_x \delta \mu$. Hence, from Eq.~(\ref{eq:w2}) we have that $w \propto ( \partial_x \delta \mu )^2$. From Eq.~(\ref{eq:dist2}), it is clear that both $\partial_x u$ and $\partial_x \delta \mu$ generate perturbations to the distribution function that are odd in momentum, thus allowing for the presence of the cross term $(\partial_x u)(\partial_x \delta \mu)$. Thus, the dissipated power must have the form
\begin{equation}
w = \zeta \left( \partial_x u \right)^2 + \gamma \left( \partial_x \delta \mu \right)^2 + 2 \lambda \left( \partial_x u \right) \left( \partial_x \delta \mu \right),
\label{eq:w4}
\end{equation}
where we have introduced two additional transport coefficients, $\gamma$ and $\lambda$. In order for $w$ to be nonnegative, these coefficients must satisfy $\zeta, \gamma > 0$ and $\lambda^2 \leq \zeta \gamma$.

The appearance of additional transport coefficients $\gamma$ and $\lambda$ at finite frequencies is a result of the breakdown of the single fluid description of the 1D Fermi gas. Indeed, it was shown recently~\cite{matveev_second_2017,matveev_propagation_2018,matveev_propagation_2018,matveev_two-fluid_2019} that at low temperatures 1D Fermi systems are described by two-fluid hydrodynamics analogous to that of superfluid He-4~\cite{landau_fluid_1987,khalatnikov_introduction_2000}. Given this correspondence, it is instructive to introduce parallel notation. For the case of superfluid He-4, the viscous coefficients are defined via the mass current~\cite{khalatnikov_introduction_2000}. For a generic dispersion, however, mass current is not a meaningful quantity. Instead, we can express the dissipated power in terms of the particle number current $j_n$, which is defined by
\begin{equation}
j_n = \frac{\nu}{h} \int_{-\infty}^\infty v_p n_p \, dp ,
\label{eq:jn}
\end{equation}
where $v_p = \partial \varepsilon_p / \partial p$. For $T \ll E_F$ and $u \ll v_F$, we have
\begin{equation}
j_n = n u + \frac{\nu}{h} \delta \mu,
\label{eq:jnform}
\end{equation}
where $n$ is the particle density. This expression is obtained by substituting the distribution function (\ref{eq:dist2}) into the definition (\ref{eq:jn}). Using Eq.~(\ref{eq:jnform}), we express $\delta \mu$ in terms of $j_n$ and $u$, thus bringing Eq.~(\ref{eq:w4}) to the form
\begin{eqnarray}
w = \zeta_2 \left( \partial_x u \right)^2 + \zeta_3 \left[ \partial_x \left( j_n - n u \right) \right]^2 \nonumber \\ + 2 \zeta_1 \left[ \partial_x \left( j_n - n u \right) \right] \left( \partial_x u \right),
\label{eq:w6}
\end{eqnarray}
where
\begin{eqnarray}
\zeta_1 = \frac{h}{\nu} \lambda, \quad
\zeta_2 = \zeta, \quad
\zeta_3 = \left( \frac{h}{\nu} \right)^2 \gamma.
\end{eqnarray}
Equation~(\ref{eq:w6}) is the one-dimensional analog of the well known expression for the dissipation rate in superfluid He-4~\cite{khalatnikov_introduction_2000}.

We now calculate the viscosities $\gamma$ and $\lambda$ by following the procedure described in the previous section. In particular, we consider a point in the gas at which both $u$ and $\delta \mu$ vanish but the gradients of these quantities are non-zero. We begin by substituting the distribution function~(\ref{eq:dist2}) into Eq.~(\ref{eq:boltzmann1}). This gives
\begin{eqnarray}
\dot{n}_p = \frac{g_p^2}{T} \left[ \frac{\varepsilon_p - \mu}{T} \partial_t T + \partial_t \mu + p \partial_p \varepsilon_p \partial_x u \right. \nonumber \\ \left. + \frac{1}{2} \textrm{sgn}(p) \partial_p \varepsilon_p \partial_x  \delta \mu \right].
\label{eq:leftboltz5}
\end{eqnarray}
The expression for $\dot{n}_p$ may be cast in the form of Eq.~(\ref{eq:leftboltz4}), where now
\begin{equation}
\Upsilon(\xi) = \partial_t \mu + \frac{\partial_t T}{T} \xi +  \frac{p(\mu + \xi)}{p'(\mu + \xi)} \partial_x u +  \frac{1}{2 p'(\mu + \xi)} \partial_x \delta \mu.
\end{equation}
Applying conservation of energy and particle number (\ref{eq:conservation}), we obtain two linear relations involving $\partial_t T$, $\partial_t \mu$, $\partial_x u$, and $\partial_x \delta \mu$. The quantities $\partial_t T$ and $\partial_t \mu$ may thus be eliminated from Eq.~(\ref{eq:leftboltz4}) in favor of $\partial_x u$ and $\partial_x \delta \mu$. To leading order in $T/E_F$, this procedure yields $\dot{n}_p$ of the form (\ref{eq:np}) where $\phi_p$ is still given by Eq.~(\ref{eq:phi}) but the expression (\ref{eq:upsilonzero}) is replaced by
\begin{equation}
\Upsilon''(0) = - \chi^{\phantom\dagger}_0  \partial_x u - \eta^{\phantom\dagger}_0 \partial_x \delta \mu.
\label{eq:upsilonzero2}
\end{equation}
Here the quantity $\chi^{\phantom\dagger}_0$ is again given by Eq.~(\ref{eq:chi}) and
\begin{equation}
\eta^{\phantom\dagger}_0 = - \left. \left( \frac{1}{2 p'(\varepsilon)} \right)''  \right \vert_{\varepsilon = \mu} = - \left. \frac{1}{2\varepsilon_p'} \left( \frac{\varepsilon_p''}{ \varepsilon_p'} \right)' \right\vert_{p = p_F}.
\label{eq:chitilde}
\end{equation}

We now apply Eqs.~(\ref{eq:w2}), (\ref{eq:np}), and (\ref{eq:upsilonzero2}) to obtain
\begin{equation}
w = \frac{1}{4 T} \left( \chi^{\phantom\dagger}_0 \partial_x u + \eta^{\phantom\dagger}_0 \partial_x \delta \mu \right)^2 \langle \phi | \hat{\Gamma}^{-1} | \phi \rangle.
\label{eq:w5}
\end{equation}
The inner product appearing in this formula can be expressed in terms of $\tau$ using its definition (\ref{eq:tau}). Matching the two forms of $w$ given by Eqs.~(\ref{eq:w4}) and (\ref{eq:w5}), we recover Eq.~(\ref{eq:zeta}) for $\zeta$ and obtain
\begin{eqnarray}
\gamma = \left(\frac{\eta^{\phantom\dagger}_0}{\chi^{\phantom\dagger}_0} \right)^2 \zeta, \quad
\lambda = \frac{\eta^{\phantom\dagger}_0}{\chi^{\phantom\dagger}_0} \zeta. \label{eq:gammalambda}
\end{eqnarray}
We observe that these values of $\gamma$ and $\lambda$ saturate the inequality $\lambda^2 \leq \zeta \gamma$, which is a feature of working to only leading order in $T/E_F$.

Equation~(\ref{eq:special_dispersion1}) gives the general form of the dispersion for which $\chi^{\phantom\dagger}_0$ vanishes at any density. We now derive the analog for $\eta_0$. Setting the right-hand side of Eq.~(\ref{eq:chitilde}) to zero and solving the resultant differential equation, we obtain~\cite{Note3}
\begin{equation}
\varepsilon_p = A \exp\left( B \left| p \right| \right) + C.
\label{eq:exp}
\end{equation}
It is worth mentioning that for $B = c/A$ and $C = - A$ in the limit that $A$ tends to infinity,  Eq.~(\ref{eq:exp}) reduces to the ultrarelativistic dispersion $\varepsilon_p = c | p |$. That $\eta^{\phantom\dagger}_0$ vanishes in this case is apparent from Eq.~(\ref{eq:chitilde}) given the presence of $\varepsilon_p''$ in the second expression for $\eta^{\phantom\dagger}_0$. The physics underlying the vanishing of $\eta^{\phantom\dagger}_0$ is discussed in Sec.~\ref{sec:scaleinvariance}.

\section{Weak Interactions}
\label{sec:interactions}

So far, our consideration of interactions has focused exclusively on their role in restoring the gas to equilibrium. These effects enter the expression for the bulk viscosity (\ref{eq:zeta}) through the relaxation time $\tau$ defined by Eq.~(\ref{eq:tau}). However, interactions also alter the effective dispersion $\varepsilon_p$ appearing in Eq.~(\ref{eq:boltzmann1}), cf. Ref.~\cite{degottardi_viscous_2020}. While the resultant correction is small for weak interactions, this becomes a crucial consideration if lattice effects are also weak. In Sec.~\ref{sec:interactionsA}, we calculate the bulk viscosity $\zeta$ of a weakly interacting gas. Then, in Sec.~\ref{sec:interactionsB}, we study the competition between lattice effects and interactions in the context of the tight-binding model.

\subsection{Effect of interactions on the bulk viscosity}

\label{sec:interactionsA}

We consider a weak two-particle interaction described by the Hamiltonian
\begin{equation}
  \label{eq:interaction_hamiltonian}
  \hat{V} = \frac{1}{2L}\sum_{\substack{p,p',q\\ \sigma,\sigma'}}
  V(q)a_{p+q,\sigma}^\dagger a_{p'-q,\sigma'}^\dagger
      a_{p',\sigma'}^{}a_{p,\sigma}^{}.
\end{equation}
Here, $V(q)$ is the Fourier transform of the interaction potential, and
$a_{p,\sigma}^{}$ annihilates a fermion of momentum $p$ and $z$-component of spin $\sigma$. To first order, the energy of the state with occupation numbers $n_{p\sigma}$ is
\begin{equation}
  \label{eq:E}
  E=\sum_{p,\sigma} \varepsilon_p n_{p\sigma}
  +\frac{1}{2L}\sum_{\substack{p,p'\\ \sigma,\sigma'}}
  [V(0)-V(p-p')\delta_{\sigma,\sigma'}]n_{p\sigma}n_{p'\sigma'}.
\end{equation}
For spinless systems, this expression is applicable as long as $V(0)/\hbar v_F \ll 1$. For systems with spin, spin-charge separation \cite{dzyaloshinskii_correlation_1974,giamarchi_quantum_2004} would seem to preclude the application of perturbation theory. Fortunately however, such effects are negligible for $p_F V(0)/\hbar \ll T$~\cite{matveev_relaxation_2020,karzig_energy_2010}. Given that $E$ in Eq.~(\ref{eq:E}) has the form of the Fermi liquid expression for interactions between quasiparticles~\cite{lifshitz_statistical_1980}, we may apply the well established procedure of calculating transport coefficients in Fermi liquid theory~\cite{abrikosov_theory_1959,lifshitz_physical_1981,sykes_transport_1970} as long as we only work to first order in interactions.

In Fermi liquid theory, the quasiparticle energies are given by $\mathcal{E}_{p\sigma} = \delta E/\delta n_{p\sigma}$~\cite{lifshitz_statistical_1980}. Evaluating this quantity using Eq.~(\ref{eq:E}) gives an effective dispersion
\begin{equation}
\mathcal{E}_p = \varepsilon_p + \delta \varepsilon_p,
\label{eq:quasiparticle_energies1}
\end{equation}
where
\begin{equation}
\delta \varepsilon_p = \int \frac{dp'}{h}[\nu V(0)-V(p-p')] n_{p'}^{(0)},
\label{eq:quasiparticle_energies2}
\end{equation}
and the equilibrium distribution function now takes the form~\cite{abrikosov_theory_1959,sykes_transport_1970}
\begin{equation}
n_p^{(0)} = \frac{1}{e^{(\mathcal{E}_p - u p - \mu)/T}+1}.
\end{equation}
In deriving Eq.~(\ref{eq:quasiparticle_energies2}), we have assumed spin degeneracy and summed over spins.

We repeat our calculation of the viscosity given in Sec.~\ref{sec:boltzmann}, taking into account the correction to the dispersion arising from interactions. We work to first order in the interaction strength. In proceeding, care must be taken because $\mathcal{E}_p$ given by Eqs.~(\ref{eq:quasiparticle_energies1}) and (\ref{eq:quasiparticle_energies2}) depends on $u(x)$, $T(t)$, and $\mu(t)$ via $n_p^{(0)}$. We find~\footnote{Because $\mathcal{E}_p$ is defined through $n_p^{(0)}$, it is also a function of $u(x)$. This introduces two additional contributions
to $\dot{n}_p$ (not shown in the text) that cancel each other. These terms arise from the second and third terms on the right-hand side of Eq.~(\ref{eq:boltzmann1}).}
\begin{eqnarray}
\dot{n}_p  &=&  n_p^{(0)} \left( 1 - n_p^{(0)} \right) \frac{1}{T} \Bigg[ \left(1 - \frac{\partial \mathcal{E}_p}{\partial \mu} \right) \partial_t \mu \nonumber \\
 && + \left(  \frac{\mathcal{E}_p - \mu}{T}  - \frac{\partial  \mathcal{E}_p }{\partial T}  \right) \partial_t T + p \frac{\partial \mathcal{E}_p}{\partial p} \partial_x u \Bigg].
\label{eq:boltz1}
\end{eqnarray}

Introducing $\xi = \mathcal{E}_p - \mu$ and eliminating $p$ in favor of $\xi$, we can
cast the quantity $\dot{n}_p$ in form of Eq.~(\ref{eq:leftboltz4}) where
\begin{eqnarray}
\Upsilon(\xi) &=& \left( 1 - \frac{\partial \delta \varepsilon_{p(\mu + \xi)}}{\partial \mu} \right) \partial_t \mu + \left[ \frac{\xi}{T} - \frac{\partial \delta \varepsilon_{p(\mu + \xi)} }{\partial T}  \right] \partial_t T \nonumber \\  && + \frac{p(\mu + \xi)}{p'(\mu + \xi)} \partial_x u.
\label{eq:upsilon3}
\end{eqnarray}
Here $p(\mathcal{E})$ is the inverse function of $\mathcal{E}_p$. The terms $\partial \delta \varepsilon_{p(\mu + \xi)}/ \partial \mu$ and $\partial \delta \varepsilon_{p(\mu + \xi)} / \partial T$ vanish in the non-interacting limit. In evaluating these terms, the corrections that arise from the dependence of $p(\mu+\xi)$ on interactions enter the calculation at second order in interaction strength and thus can be neglected. In contrast, the dependence of $p(\mu + \xi)$ on interactions must be included in the final term of Eq.~(\ref{eq:upsilon3}).

We now repeat the steps leading to the expression (\ref{eq:zeta}) for the bulk viscosity. Conservation of momentum and energy gives two relations involving the infinitesimal quantities $\partial_t T$, $\partial_t \mu$, and $\partial_x u$. Eliminating $\partial_t T$ and $\partial_t \mu$ and working to leading order in $T \sim \xi \ll E_F$, we obtain Eq.~(\ref{eq:np}) with
\begin{equation}
\Upsilon''(0) = - ( \chi^{\phantom\dagger}_0 + \chi^{\phantom\dagger}_1 ) \partial_x u,
\label{eq:upsilonzero3}
\end{equation}
where $\chi^{\phantom\dagger}_0$ is given by Eq.~(\ref{eq:chi}) and
\begin{widetext}
\begin{eqnarray}
\chi^{\phantom\dagger}_1 &=& -\frac{1}{2 \pi \hbar \left( \varepsilon'(p_F) \right)^4} \bigg\{ \left[ 3 p_F \left( \varepsilon''(p_F)\right)^2 - 2 \varepsilon'(p_F) \varepsilon''(p_F) - 2 p_F \varepsilon'(p_F) \varepsilon'''(p_F) \right] \left( V(0) - V(2p_F) \right) \nonumber \\ && + \left[ 3 p_F \varepsilon'(p_F) \varepsilon''(p_F) - \left( \varepsilon'(p_F) \right)^2 \right] V'(2p_F) - 2 p_F \left(\varepsilon'(p_F) \right)^2 V''(2p_F) \bigg\}.
\label{eq:eta}
\end{eqnarray}
\end{widetext}
While the quantity $\chi_1$ receives a contribution from $\partial \delta \varepsilon_{p(\mu + \xi)}/ \partial \mu$, the term $\partial \delta \varepsilon_{p(\mu + \xi)}/ \partial T$ generates terms that are higher order in temperature. Applying Eqs.~(\ref{eq:w2}), (\ref{eq:np}), and (\ref{eq:upsilonzero3}), we find
\begin{equation}
\zeta = \frac{4 \pi^3 \nu (\chi^{\phantom\dagger}_0 + \chi^{\phantom\dagger}_1 )^2 T^4 \tau}{45 \hbar v_F}.
\label{eq:zetaeta}
\end{equation}
In the limit of weak interactions, $\chi^{\phantom\dagger}_1 \rightarrow 0$, we recover the result (\ref{eq:zeta}).

In Sec.~\ref{sec:boltzmannA}, it was found that the parameter $\chi^{\phantom\dagger}_0$ vanishes for a dispersion of the form (\ref{eq:special_dispersion1}) for any positive exponent $B$. For such dispersions, the parameter $\chi_1$ will also vanish for potentials of the form $V(p) \propto |p|^{B-1}$. We defer a discussion of this to Sec.~\ref{sec:scaleinvariance}. It is worth pointing out that $\chi^{\phantom\dagger}_1$ also vanishes for a potential $V(p)$ that is independent of $p$, which in real space corresponds to a delta function interaction potential. For such a potential, the correction $\delta \varepsilon_p$ given by Eq.~(\ref{eq:quasiparticle_energies2}) is a constant and thus represents a trivial shift of the energy $\mathcal{E}_p$.

Our expression (\ref{eq:zetaeta}) for the bulk viscosity is consistent with other results that appear in the literature. In Ref.~\cite{degottardi_viscous_2020}, the bulk viscosity of a gas of spin-$1/2$ fermions with quadratic dispersion was derived. Our expression (\ref{eq:zetaeta}) reproduces the bulk viscosity given in Ref.~\cite{degottardi_viscous_2020} for
$\varepsilon_p = p^2/2m$ and $\nu = 2$. The viscosity of a liquid of spinless fermions was studied for arbitrary interaction strength in Ref.~\cite{matveev_viscous_2017}. We find that to first order in the interaction strength, the results of that work are consistent with those presented here.

\subsection{Competition between interactions and lattice effects}

\label{sec:interactionsB}

In the regime of weak interactions, it is natural to expect that $\chi_1 \ll \chi_0$, i.e., that lattice effects dominate. However, in certain cases of experimental relevance, the dispersion can be nearly quadratic. As discussed in Sec.~\ref{sec:tight-binding}, in such cases $\chi_0$ tends to zero, and $\chi_1$ may be expected to become the dominant contribution to the bulk viscosity (\ref{eq:zetaeta}).

We thus explore the competition between interactions and lattice effects. As an example, we consider the tight-binding model at low fermion density, $n a \ll 1$, so that the dispersion approaches the quadratic form (\ref{eq:mstar}). We further assume that the fermions are spinless ($\nu = 1$) and discuss separately the cases of short- and long-range potentials. For a short-range potential, we have~\footnote{Here, a potential is considered short-range if the small $q$ expansion (\ref{eq:expansion}) applies.}
\begin{equation}
V(q) = V(0) + \frac{1}{2} V''(0) q^2,
\label{eq:expansion}
\end{equation}
where we have assumed that $V(x)$ decays faster than $1/|x|^3$. In this case, the dimensionless interaction strength is
\begin{equation}
\frac{V(0) - V(2p_F)}{\pi \hbar v_F} = - 2 m^\ast n V''(0),
\label{eq:pertcond}
\end{equation}
where the effective mass $m^\ast$ is defined by Eq.~(\ref{eq:mstar}) and the particle density $n = p_F/\pi \hbar$ for $\nu = 1$. The applicability of the weak interaction approximation of Sec.~\ref{sec:interactionsA} requires that the parameter (\ref{eq:pertcond}) be much less than unity. To leading order in $n a \ll 1$, the ratio of $\chi_1$ [Eq.~(\ref{eq:chi})] to $\chi_0$ [Eq.~(\ref{eq:eta})] is given by
\begin{equation}
\frac{\chi_1}{\chi_0} = \frac{5}{2} m^\ast n V''(0).
\label{eq:chi1overchi0}
\end{equation}
We find that the ratio $\chi_1/\chi_0$ is of the same order of magnitude as the small parameter (\ref{eq:pertcond}). Thus, for weak short-range interactions, $\chi_1$ is a small correction to $\chi_0$.

We now consider long-range interactions. As an example, we take
\begin{equation}
V(x) = \frac{e^2}{|x|} - \frac{e^2}{\sqrt{x^2 + 4 d^2}}.
\end{equation}
This is the Coulomb interaction screened at large distances by a gate modeled as a metal plane at a distance $d$ from the system. At $n d \gg 1$, the small parameter of the perturbation theory is given by
\begin{equation}
\frac{V(0) - V(2p_F)}{\pi \hbar v_F} = \frac{2}{\pi^2 n a_B} \log \left( 2 \pi n d  \right),
\label{eq:pertcond2}
\end{equation}
where $a_B = \hbar^2/m^\ast e^2$ is the Bohr radius. Neglecting the logarithmic factor, we conclude that perturbation theory holds as long as $n a_B \gg 1$. From Eqs.~(\ref{eq:chi}) and (\ref{eq:eta}), we have
\begin{equation}
\frac{\chi_1}{\chi_0} = - \frac{3}{2 \pi^4 n^3 a_B a^2} \log \left( 2 \pi n d  \right).
\label{eq:eta1vseata02}
\end{equation}
Neglecting the logarithm, we find the condition that this ratio greatly exceeds unity can be written as $n^3 a_B a^2 \ll 1$. In summary, the condition $| \chi_1 / \chi_0 | \gg 1$ holds in the perturbative regime provided that
\begin{equation}
\frac{1}{a_B} \ll n \ll \frac{1}{a_B^{1/3} a^{2/3}}.
\label{eq:range}
\end{equation}
For sufficiently weak interactions we have $a_B \gg a$, which guarantees the existence of the range (\ref{eq:range}).

We conclude that for weak interactions, there are scenarios in which either lattice effects or interactions may dominate the viscous behavior of a Fermi gas. As we have seen, this behavior is closely linked to whether the interactions are short- or long-range.

\section{Suppression of Viscosity}
\label{sec:scaleinvariance}

For dispersions of the form~(\ref{eq:special_dispersion1}), the parameter $\chi_0$ vanishes, signalling the suppression of $\zeta$. For the special case of a quadratic dispersion ($B = 2$), this is the analog of the well known result that the bulk viscosity of a three-dimensional classical gas is suppressed~\cite{lifshitz_physical_1981} due to the scale invariance of the fermion dispersion~\cite{matveev_viscous_2017,degottardi_viscous_2020}. In this section, we show that the same argument explains the vanishing of $\chi_0$ for the power-law dispersion~(\ref{eq:special_dispersion1}) with any exponent $B$. Similarly, the parameter $\eta_0$, which appears in the expressions for the viscosities $\gamma$ and $\lambda$ given by Eq.~(\ref{eq:gammalambda}), vanishes for dispersions of the form (\ref{eq:exp}). We will show that this behavior can be explained with similar arguments.

We begin by considering the case of the bulk viscosity $\zeta$ in the case of the power-law dispersion~(\ref{eq:special_dispersion1}), where without loss of generality we can set $C=0$. As in Sec.~\ref{sec:boltzmann}, we consider a gas, initially in thermal equilibrium, subject to an infinitesimal velocity gradient $\partial_x u$. We will show that for dispersions of the form (\ref{eq:special_dispersion1}), this perturbation will not drive the system out of equilibrium. From the continuity equation, the infinitesimal gradient of the velocity $u$ is equivalent to a time-dependent particle density, $\partial_t n = - n \partial_x u$. We consider the scenario in which the particle number $N$ is fixed but the length of the system $L$ is time-dependent. Finite size quantization requires that the momentum of any state $p$ be quantized in units of $\pi\hbar/L$. For a power-law dispersion of the form~(\ref{eq:special_dispersion1}), $\varepsilon_p$ therefore scales as $1/L^B$. To wit, as the system size changes from $L(0) = L_0$ to $L(t) = L$, the energy of a state of momentum $p$ evolves from $\varepsilon_p^{(0)}$ to
\begin{equation}
\varepsilon_{p(t)} = \varepsilon_p^{(0)}\left( \frac{L_0}{L} \right)^{B}.
\label{eq:scaling2}
\end{equation}
Let us assume that at $t = 0$ the system is in equilibrium, and the occupation numbers of the fermion states are given by the Fermi-Dirac distribution
\begin{equation}
n_p = \frac{1}{e^{(\varepsilon_p - \mu)/T}+1}
\label{eq:FD}
\end{equation}
with $\varepsilon_p = \varepsilon_p^{(0)}$. A generic change to $\varepsilon_p$ would violate the relation (\ref{eq:FD}) and thus would drive the system out of equilibrium. However, in the case of the power-law dispersion, $\varepsilon_p$ changes by a factor that does not depend on $p$. As a result, by choosing new values of the temperature and chemical potential according to
\begin{equation}
T = T_0 \left( \frac{L_0}{L} \right)^{B}
\label{eq:Tchange}
\end{equation}
and
\begin{equation}
\mu = \mu_0 \left( \frac{L_0}{L} \right)^{B},
\end{equation}
we find that the distribution function $n_p$ retains its Fermi-Dirac form (\ref{eq:FD}). Thus, the system remains in equilibrium, and the rate $\dot{n}_p$ must vanish.

The above argument applies only in the limit of weak interactions. We now show that for certain types of interactions, the bulk viscosity vanishes regardless of the interaction strength. Consider a Hamiltonian with kinetic energy described by $\varepsilon_p \propto |p|^B$ and interactions given by Eq.~(\ref{eq:interaction_hamiltonian}) with $V(q) \propto |q|^{B-1}$. A system in thermal equilibrium is described by the Gibbs distribution
\begin{equation}
w_n = \frac{1}{Z} e^{-E_n / T},
\label{eq:Gibbs}
\end{equation}
where $w_n$ is the probability of finding the system in a state with energy $E_n$, and $Z$ is the partition function~\cite{landau_statistical_2013}. As above, we take the system to have a time-dependent length $L(t)$ and consider the scaling of the energy $E_n$. The Hamiltonian is composed of an operator describing kinetic energy, which scales as $1/L^B$ in accordance with Eq.~(\ref{eq:scaling2}), and an operator describing interactions. Noting the factor of $1/L$ in Eq.~(\ref{eq:interaction_hamiltonian}), the interaction Hamiltonian also scales as $1/L^B$ provided that $V(q) \propto |q|^{B-1}$. As a result, the total Hamiltonian scales as $1/L^B$ and thus its eigenvalues $E_n$ must also scale as $1/L^B$. If the system starts in equilibrium, then it will remain in equilibrium with a distribution described by Eq.~(\ref{eq:Gibbs}) and the time-dependent temperature given by Eq.~(\ref{eq:Tchange}). As we saw in Sec.~\ref{sec:interactions}, the linear in interactions contribution to $\chi$, given by $\chi_1$, does indeed vanish in this case.  The above argument is more general and shows that the viscosity must vanish in all orders in the interaction strength.

We now consider a system with a dispersion of the form (\ref{eq:exp}) and demonstrate that an infinitesimal gradient of $\delta \mu$ does not drive the system out of equilibrium. We recall that a system with a spatially uniform $\delta \mu$ has an equilibrium distribution (\ref{eq:dist2}), in which we set $u = 0$. This distribution can be formally interpreted as the standard Fermi-Dirac distribution (\ref{eq:FD}) for particles with energies $\varepsilon_{p} \to \varepsilon_p + U(p)$,
where $U(p)$ is the momentum-dependent potential
\begin{equation}
U(p) = - \frac{1}{2} \textrm{sgn}(p) \delta \mu.
\label{eq:potential}
\end{equation}
We now take $\delta \mu$ to have an infinitesimal gradient. This gives rise to a momentum dependent force
\begin{equation}
 - \frac{\partial U}{\partial x} = \textrm{sgn}(p) \frac{\partial_x \! \left( \delta \mu \right)}{2}
\end{equation}
acting on a particle with momentum $p$. As a result, the momentum of each particle evolves in time according to
\begin{equation}
p(t) = p + \textrm{sgn}(p) \frac{\partial_x \! \left( \delta \mu \right)}{2} t.
\end{equation}
For the dispersion given by Eq.~(\ref{eq:exp}), the energy of a particular state is given by
\begin{equation}
\varepsilon_{p(t)} = \varepsilon_p^{(0)} e^{B (\partial_x \delta \mu) t/2}.
\end{equation}
[We have set $C$ in Eq.~(\ref{eq:exp}) equal to zero]. Generically, a change in $\varepsilon_p$ drives the system out of equilibrium. But as was the case in Eq.~(\ref{eq:scaling2}), the energies $\varepsilon_{p(t)}$ change by a $p$-independent factor. As long as the temperature and chemical potential have the same time dependences, namely
\begin{equation}
T = T_0 e^{B (\partial_x \delta \mu) t/2}, \quad \mu = \mu_0 e^{B (\partial_x \delta \mu) t/2},
\end{equation}
the system is described by the equilibrium distribution (\ref{eq:FD}). We conclude that an infinitesimal $\partial_x \delta \mu$ does not drive a system described by the dispersion (\ref{eq:exp}) out of equilibrium, consistent with the fact that $\eta_0 = 0$.

\section{Discussion and Results}

\label{sec:discconcl}

We have presented a systematic study of the bulk viscosity for one-dimensional Fermi gases with arbitrary dispersions, and thus the theory can account for lattice effects. The expression for the bulk viscosity, given by Eq.~(\ref{eq:zetaeta}), has the general form
\begin{equation}
\zeta \propto \chi^2 \tau,
\label{eq:zeta2}
\end{equation}
i.e., the bulk viscosity is controlled by two quantities: the relaxation time $\tau$ and the parameter $\chi$. The latter is a measure of the \emph{sensitivity} of the gas to the velocity gradient $\partial_x u$ and quantifies the extent to which this perturbation displaces the gas from equilibrium. For instance, we found that a gas of free fermions with a dispersion given by Eq.~(\ref{eq:special_dispersion1}) is insensitive to gradients of $u$, resulting in $\chi_0 = 0$. (In the absence of interactions, $\chi = \chi_0$.)

To appreciate the central role played by $\chi$, we consider a gas of fermions with a dispersion given by Eq.~(\ref{eq:special_dispersion1}), for which $\chi$ vanishes in the limit of weak interactions due to scale invariance. Given the expression (\ref{eq:zeta2}), we expect that $\zeta$ vanishes in this limit. However, this conclusion is premature since interactions also control the relaxation properties of the system. In the limit that interactions vanish, $\tau$ diverges, and thus the expression (\ref{eq:zeta2}) for the bulk viscosity is indeterminate. To determine the fate of $\zeta$, we must therefore consider the regime of weak but non-vanishing interactions~\cite{matveev_viscous_2017}. While $\chi_0$ vanishes, $\chi_1$, given by Eq.~(\ref{eq:eta}), is proportional to the interaction strength $V$, and thus $\chi \propto V$. On the other hand, the relaxation in one dimension is dominated by three-particle processes, for which $\tau \propto 1/V^4$~\cite{lunde_three-particle_2007,khodas_fermi-luttinger_2007}. Thus, in the limit of weak interactions, the bulk viscosity (\ref{eq:zeta2}) is still large $\zeta \propto 1/V^2$, but is suppressed compared with that of Fermi gases with generic dispersions, for which $\zeta \propto 1/V^4$.

The conclusion that interactions in a Fermi gas with power-law dispersion (\ref{eq:special_dispersion1}) result in a nonvanishing viscosity assumes that the interaction will spoil the scale invariance of the system. However, for an interaction that satisfies $V(q) \propto |q|^{B-1}$ where $B$ is the exponent in Eq.~(\ref{eq:special_dispersion1}), the exact many-body energy levels scale as a power of the system size. As a result, the bulk viscosity vanishes regardless of the strength of the interactions. This argument is not limited to 1D systems. An example of a three-dimensional system with zero bulk viscosity is the Fermi gas in the unitary limit~\cite{son_vanishing_2007}.

A peculiarity of one-dimensional quantum systems is that for particular interactions they can possess an infinite number of conserved quantities. Such systems are described by \emph{integrable models}. These systems do not relax, i.e., the relaxation time $\tau$ is formally infinite, and thus the bulk viscosity (\ref{eq:zeta2}) is infinite for integrable models. An exception is the Calogero-Sutherland model~\cite{sutherland_beautiful_2004}, which describes particles with dispersion $\varepsilon_p\propto |p|^B$ and interactions $V(q)\propto |q|^{B-1}$ with $B=2$. As discussed in the previous paragraph, this model is insensitive to the velocity gradient, so $\chi=0$. On the other hand, by virtue of integrability, the relaxation time $\tau$ in Eq.~(\ref{eq:zeta2}) is infinite. The bulk viscosity (\ref{eq:zeta2}) is thus indeterminate.

We also considered the case of Fermi gases driven at finite frequencies. For a broad range of frequencies, these systems fail to come to full equilibrium and are instead described by the distribution (\ref{eq:dist2}) in which the parameter $\delta \mu$ is the difference between the chemical potentials of right and left movers. Position dependence of this parameter, $\partial_x \delta \mu \neq 0$, leads to viscous dissipation, which is described by the quadratic form~(\ref{eq:w4}). This expression defines two additional bulk viscosities, $\gamma$ and $\lambda$. As discussed above, systems of particles with dispersion (\ref{eq:special_dispersion1}) are insensitive to the perturbation $\partial_x u$ in that it does not drive the system out of equilibrium. Similarly, a gas of fermions with dispersion (\ref{eq:exp}) is insensitive to the perturbation $\partial_x \delta \mu$. It is worth noting that gases of fermions obeying the ultra-relativistic dispersion $\varepsilon_p = c|p|$, which is a special case of both Eqs.~(\ref{eq:special_dispersion1}) and (\ref{eq:exp}), are  insensitive to both gradients of $u$ and $\delta \mu$.

Finally, given the importance of interactions to viscous properties, it is natural to ask whether the results of this work can be extended beyond the weakly interacting limit. In fact, viscous dissipation of a spinless Luttinger liquid was considered in Ref.~\cite{matveev_viscous_2017}, where the interaction strength was not assumed to be weak. Though the focus of that work was on Galilean invariant systems, much of the discussion applies to arbitrary dispersions. Unfortunately, making a similar generalization to spinful systems is not straightforward given that their relaxation properties are not well understood.

\begin{acknowledgments}
The authors are grateful to A. V. Andreev for stimulating discussions. Work at Argonne National Laboratory was supported by the U.S. Department of Energy, Office of Science, Basic
Energy Sciences, Materials Sciences and Engineering Division.
\end{acknowledgments}

\bibliography{library}

\end{document}